# Electric field modulation of Schottky barrier height in graphene/MoSe$_2$ van der Waals heterointerface

Yohta Sata[1], Rai Moriya[1,*], Sei Morikawa[1], Naoto Yabuki[1], Satoru Masubuchi[1,2], and Tomoki Machida[1,2,*]

[1] *Institute of Industrial Science, University of Tokyo, 4-6-1 Komaba, Meguro, Tokyo 153-8505, Japan*

[2] *Institute for Nano Quantum Information Electronics, University of Tokyo, 4-6-1 Komaba, Meguro, Tokyo 153-8505, Japan*

We demonstrate a vertical field-effect transistor based on a graphene/MoSe$_2$ van der Waals (vdW) heterostructure. The vdW interface between the graphene and MoSe$_2$ exhibits a Schottky barrier with an ideality factor of around 1.3, suggesting a high-quality interface. Owing to the low density of states in graphene, the position of the Fermi level in the graphene can be strongly modulated by an external electric field. Therefore, the Schottky barrier height at the graphene/MoSe$_2$ vdW interface is also modulated. We demonstrate a large current ON-OFF ratio of $10^5$. These results point to the potential high performance of the graphene/MoSe$_2$ vdW heterostructure for electronics applications.

*E-mail: moriyar@iis.u-tokyo.ac.jp; tmachida@iis.u-tokyo.ac.jp

Heterostructures that are held together by the van der Waals (vdW) force between the layered materials have been the subject of considerable interest in the fields of materials science and opto-electronics [1]. To date, several layered materials have been developed and extensively studied. Such materials include graphene, black phosphorous, transition metal dichalcogenides (TMD), and hexagonal boron nitride. Among these materials, heterostructures based on graphene and TMD have been found to exhibit functions that were previously not possible, including those of a vertical field-effect transistor [2-5], photocurrent generation [6,7], a spin-orbit proximity effect [8], and the ability to fabricate flexible devices [9]. Particularly, vertical field-effect transistors based on graphene/$MoS_2$/metal heterostructures have been the subject of considerable attention mainly due to their large current ON-OFF ratio ($10^3$ to $10^5$) combined with a large ON current density (>$10^3$ A/cm$^2$) [3,4]. This level of performance would attract very high demand for electronics applications. Given that the large current ON-OFF ratio is a result of the electric field modulation of the Schottky barrier height at the graphene/$MoS_2$ interface, the precise tuning of the band alignment at the graphene/TMD interface is crucial in determining the level of performance. To date, such devices have been fabricated using $MoS_2$ which has an indirect band gap of around 1.3 eV in its bulk form [10]. Changing the chalcogen atom from sulfur to selenium significantly reduces the band gap (the indirect gap of $MoSe_2$ is around 1.1 eV) [11] and also changes the electron affinity [12,13]. Furthermore, $MoSe_2$ exhibits better optical properties than $MoS_2$ [11,14]. These results point to $MoSe_2$ being the better option for opto-electronics applications. In addition, since $MoSe_2$ has a smaller electron affinity than $MoS_2$, under the assumption of Schottky–Mott rule, the Schottky barrier height at graphene/$MoSe_2$ interface is expected to be larger than that of graphene/MoS$_2$. Therefore a comparison of the Schottky barrier property of different TMD materials provides an insight into the vdW interface properties of a graphene/TMD heterostructure. In this study, we fabricated a graphene/MoSe$_2$/Ti vertical field-effect transistor with a graphene/MoSe$_2$ vdW interface and studied its gate-tunable properties.

The device structure of the vertical field-effect transistor is illustrated in Fig. 1(a). By mechanically exfoliating the Kish graphite, monolayer flakes of graphene are fabricated on a 300-nm SiO$_2$/$n^+$-Si substrate. Next, MoSe$_2$ flakes are mechanically exfoliated from a bulk crystal (HQ Graphene Inc.) and then transferred to the graphene by using the dry transfer method [15,16] to form a vdW junction between the freshly cleaved surface of the graphene and MoSe$_2$. Finally, by using standard electron beam (EB) lithography and EB evaporation, 30-nm Au/50-nm Ti electrodes are fabricated on both the graphene and MoSe$_2$. We fabricated a series of devices with different thicknesses of MoSe$_2$, ranging from 8.9 nm to 52.6 nm, confirming these thicknesses by using atomic force microscope (AFM) measurements. These thicknesses correspond to a layer number range of $N$ = 14 to 81, assuming the thickness $d$ of a single monolayer of MoSe$_2$ to be 0.65 nm. The junction area measured 0.3 to 1.7 µm$^2$. The fabricated device with 44 layers of MoSe$_2$ ($N$ = 44) is shown in Fig. 1(b). The electrical transport was measured in a variable-temperature cryostat. For this, the bias voltage $V_B$ was applied between the metal and graphene to drive the current in the vertical heterostructure. The back gate voltage $V_G$ was applied to a highly doped Si substrate to control the carrier concentration in the graphene.

The band alignment of the graphene/MoSe$_2$/Ti structure is shown in Fig. 1(c). The application of a negative $V_G$ reduces the electron concentration in the graphene and lowers its Fermi level. Here, the Schottky barrier height at the vdW interface increases and its conductance decreases as illustrated in the left panel of Fig. 1(c). Applying a positive $V_G$ increases the Fermi level of the graphene due to the electron doping. In this case, the Schottky barrier height decreases and the conductance of the vdW interface increases. Therefore, the graphene/MoSe$_2$ vdW interface acts as a Schottky diode with a gate-tunable barrier height. Therefore, the device acts as a vertical field-effect transistor [3,4].

The measured transport data for graphene/MoSe$_2$/Ti vertical heterostructures for numbers of MoSe$_2$ layers $N$ = 14, 44, and 75 are shown in Fig. 2. In the figures, the current–voltage (*I*–*V*) characteristics measured for different values of $V_G$ ($V_G$ = +50 V and -50 V) are shown. The current values are normalized with the junction areas of each device. The *I-V* curves display significant $N$ dependence. For a device for which $N$ = 44, shown in Fig. 2(b), the *I*–*V* curve for $V_G$ = -50 V exhibits a strong current rectification similar to that of a Schottky diode, while the device also exhibits its lowest conductivity; we define this condition as the OFF state. On the other hand, the *I*–*V* curve for $V_G$ = +50 V is nearly symmetrical with respect to $V_B$ and the device exhibits its highest conductivity; we define this as the ON state. These states are consistent with an explanation based on the modulation of a Schottky barrier, as shown in Fig. 1(c). The quality of the Schottky barrier can be evaluated from the fitting of the *I*–*V* curve in the negative-$V_B$ region at $V_G$ = -50 V, as done using thermionic emission theory [17]. The fitting results are indicated by the dashed lines in Fig. 2. We determined the ideality factor *n* of the Schottky barrier at the graphene/MoSe$_2$ interface for which $N$ = 44 as *n* = 1.3. An ideality factor approaching unity suggests a high-quality vdW interface. We demonstrated a large current ON-OFF ratio of >10$^5$ for a device for which $N$ = 44, where the current ON-OFF ratio is defined as (ON-state current)/(OFF-state current) at $V_B$ = +0.5 V. This value is comparable to the highest ON-OFF ratio achieved in a graphene/MoS$_2$/Ti vertical field-effect transistor [4]. The behavior is notably different in a device using a thinner or thicker MoSe$_2$, as shown in Figs. 2(a) and 2(c), when $N$ =14 and 75, respectively. The current rectification and gate modulation are both less pronounced relative to that for which $N$ = 44. Furthermore, we observed an increase in the ideality factor for a device for which $N$ = 14; we derived $n$ = 2.4 from the fitting data at $V_G$ = -50 V in Fig. 2(a). On the other hand, we derived $n$ = 1.3 for a device forwhich $N$ = 75. These results suggest that the properties of a graphene/MoSe$_2$/Ti vertical heterostructure are highly dependent on the value of $N$.

To undertake a more detailed comparison of the MoSe$_2$ thickness dependence, the temperature dependence of the conductance was measured to determine the Schottky barrier height at the graphene/MoSe$_2$ interface. The $V_G$ dependence of the current when $V_B$ is fixed to +0.1 V, as measured between 240 K and 300 K, was measured for both $N$ = 14 and 44. The results are shown in Figs. 3(a) and (b). The conductance monotonically decreases at lower temperatures for all values of $V_G$. The value of ln($I/T^2$) was plotted against (1000/$T$), as shown in Fig. 3(c) for $V_G$ = +40 V and -40 V. The results reveal a linear relationship, implying that the transport in the heterostructure is dominated by thermionic emission across the Schottky barrier. We fitted these data points to the thermionic emission theory and the results are plotted as solid and dashed lines in the figure. The slope of the fitting line is given by $\varphi/k_B$, where $\varphi$ denotes the effective Schottky barrier height and $k_B$ is the Boltzman constant. Given the temperature dependence at different values of $V_G$, the relationship between $\varphi$ and $V_G$ was obtained, as plotted in Fig. 3(d). These results demonstrate the direct determination of the electric field modulation of the Schottky barrier at the graphene/MoSe$_2$ vdW interface. When this figure is compared with Figs. 2(a) and (b), it is evident that the conductance of the device decreases (increases) when the barrier height is high (low). The change in the value of $\varphi$ is very non-linear under the sweep of $V_G$. This is related to the change in the graphene's Fermi level with respect to $V_G$. Analytically, this is expressed as $\mu_G = \text{sign}(V_G)\hbar v_F \sqrt{\pi \alpha V_G}$, where $\hbar$ denotes Planck's constant, $v_F = 1.0 \times 10^6$ m/s, which is the Fermi velocity of graphene, and $\alpha = 7.1 \times 10^{10}$ cm$^{-2}$V$^{-1}$, which is the capacitance of the 300-nm-thick SiO$_2$ in units of electron charge. Thus, the change in $\mu_G$ with respect to that of $V_G$ is largest around the Dirac point with the change becoming smaller as $V_G$ moves further away from the Dirac point. The data on $\varphi$ versus $V_G$, as shown in Fig. 3(d), also reflects a rapid change in $\varphi$ when $V_G$ approaches the Dirac point of the graphene electrode. The $\varphi$ values tend to saturate when $V_G$ is some distance from the Dirac point. However, the obtained $\varphi$ values exhibit quite different behaviors depending on whether $N$ = 14 or 44 [Fig. 3(d)]. The value of $\varphi$ decreases and approaches a similar value for both values of $N$ in the region of $V_G$ = +50 V. We defined this value as $\varphi_{min}$. On the other hand, the value of $\varphi$ increases and approaches different values in the region of $V_G$ = -50 V. We defined this value as $\varphi_{max}$.

A series of devices with different $N$ were fabricated and the values of $\varphi_{min}$ and $\varphi_{max}$ were obtained as summarized in Fig. 4(a). The value of $\varphi_{min}$ is similar regardless of the value of $N$, whereas $\varphi_{max}$ is highly dependent on the value of $N$. In this figure we also plot the current density for the ON state $I_{ON}$ and the OFF state $I_{OFF}$, measured at $V_B$ = +0.5 V, as shown in Fig. 4(b). The current density also shows significant $N$ dependence. Particularly, the value of $I_{OFF}$ changes more significantly with $N$. On the other hand, the value of $I_{ON}$ changes very little regardless of the value of $N$. As a result, the ON-OFF ratio as defined by $I_{ON}/I_{OFF}$ exhibits significant layer number dependence, as shown in Fig. 4(c). From the figure, it can clearly be seen that there is an optimal thickness ($N$ = 44) for the MoSe$_2$ in order to obtain a large current ON-OFF ratio. Comparing with Fig. 4(b), it can be seen that this is due to $I_{OFF}$ having a minimum value at this MoSe$_2$ thickness. At $N$ = 44, we attained the largest ON-OFF ratio of $10^5$ while, on average, $I_{ON}$ exceeded $10^3$ A/cm$^2$ for every value of $N$. These values are comparable to those for graphene/MoS$_2$/Ti vertical field-effect transistors with large ON-OFF ratio [4]. The reduction in the current ON-OFF ratio in the thinner TMD layer has also been observed for the graphene/MoS$_2$/Ti vertical heterostructure [3,4,18]. This can be explained by two dominant mechanisms; (1) the image force lowering of the Schottky barrier height and (2) the quantum capacitance of the graphene [18]. Both of these contributions depend on $N$ and are illustrated in Fig. 4(d). Considering these two contributions, the actual Schottky barrier height $\varphi$ can be calculated as $\varphi = \varphi_0 - \mu_G - \varphi_{IF}$, where $\varphi_0$ denotes the band offset between the conduction band of the MoSe$_2$ and graphene for a neutral charge, and $\varphi_{IF}$ indicates the reduction in the Schottky barrier height due to the image force lowering effect. The calculation procedure was described in detail in our previous report [18]. We used the MoSe$_2$ dielectric constant of $\varepsilon_{MoSe_2}$ = 10 [19], while the Schottky barrier height for Ti/MoSe$_2$ was $\varphi_M$ = 80 meV, as determined by the in-plane transport measurement of a bulk MoSe$_2$ layer for which $N$ = 55. The value $\varphi$ is dependent on $V_G$, whereas $\varphi_M$ is not. The barrier height determines the transport in the vertical heterostructure, as given by the larger value between $\varphi$ and $\varphi_M$, and the calculated results are plotted as the dashed line in Fig. 4(a). The calculated and measured values are in good agreement as a result of using the band offset between the conduction band of the MoSe$_2$ and graphene for a neutral charge when $\varphi_0 = 0.15$ eV.

Next, the current density is calculated using the thermionic-emission model [17]:

$$I = A\exp\left(-\frac{e\varphi}{k_\mathrm{B}T}\right)\left[-\exp\left\{\frac{e(V_\mathrm{B}-R_\mathrm{s}I)}{k_\mathrm{B}T}\right\}+1\right], \qquad (1)$$

where $A$ is a constant depending on the detail of the model, $k_\mathrm{B}$ is the Boltzmann's constant, and $R_\mathrm{s}$ is the series resistance of the device. The series resistance includes the resistance of the graphene and the contact resistance of the Au/Ti electrode in contact with the graphene or MoSe$_2$. With the values of $\varphi$ calculated as described in the previous section and $R_\mathrm{s} = 10$ k$\Omega$, the current density for $V_\mathrm{G}$ = -50 and +50 V as well as the ON-OFF ratio were calculated and plotted as the dashed lines in Figs. 4 (b) and (c). Here, we varied the value of $A$ to adjust the current density for each experiment instead of using the conventional Richardson constant. The data shown in Fig. 4(b) is reproduced by the calculation when using $A = 0.024$ Acm$^{-2}$K$^{-2}$. This value is significantly smaller than the Richardson constant extracted for a conventional metal/semiconductor Schottky diode [20,21] and we believe that such a reduction is due to the low density of states of graphene electrode [22]. Note that a similar thickness dependence whereby the ON-OFF ratio is lower for a thinner TMD layer in a vertical heterostructure has been reported for graphene/MoS$_2$/metal vertical heterostructures [3,4,18]. Thus, these behaviors are normal tendencies for graphene/TMD heterostructures. On the other hand, we observed a significant drop in the ON-OFF ratio for thicker MoSe$_2$ regions such as when $N = 74$ and 80. By comparison with Fig. 4(b), this is due to the increase in the OFF-current in the thick MoSe$_2$ layer. We can speculate that, for such values of $N$, the thickness of the MoSe$_2$ is equal to or larger than the depletion width $W$. Thus, the device behavior for $N > W/d$ is qualitatively different relative to that for $N < W/d$. We estimated the depletion width of the MoSe$_2$ by applying $W \approx \sqrt{(2\varepsilon_{MoSe_2}/eN_D)\{\varphi + eV_B - (k_B T/e)\}}$ [17], where $N_D$ is equal to around $5 \times 10^{17}$ cm$^{-3}$ and denotes the MoSe$_2$ background electron concentration. The depletion width $W$ was determined to be around 41 nm (when $N$ is around 64) with $\varphi = 0.3$ eV, $V_B = 0.5$ V, and $T = 300$ K. Calculated $W$ is close to the MoSe$_2$ thickness for which we observed the reduction in the ON-OFF ratio.

Finally, we plotted the current ON-OFF ratio $I_{ON}/I_{OFF}$ with respect to the Schottky barrier height modulation $\Delta\varphi = \varphi_{max} - \varphi_{min}$, as shown in Fig. 4(e). We plotted the data for $N < 50$. From Eq. (1), the ON-OFF ratio can be roughly related to $\Delta\varphi$ such that $I_{ON}/I_{OFF} = \exp(\Delta\varphi/k_B T)$. This expression suggests that, for a semi-log scale, the ON-OFF ratio increases linearly with $\Delta\varphi$ and this is calculated for $T = 300$ K, as plotted by the dashed line in Fig. 4(e). The measured data follows this relationship without any variable parameters. Therefore, it would appear that the transport in the graphene/MoSe$_2$/Ti vertical heterostructures is dominated by the thermionic emission at the gate-tunable graphene/MoSe$_2$ Schottky barrier, while the ON-OFF ratio is determined by $\Delta\varphi$ under the sweep of $V_G$.

In summary, we demonstrated a graphene/MoSe$_2$/Ti vertical heterostructure with a large current ON-OFF ratio of $10^5$ and an ON-current density $>10^3$ A/cm$^2$. Our results suggest the possible extension of the graphene/TMD vertical heterostructure through the use of different TMD materials with different band gaps and dielectric properties and, at the same time, having a large current modulation and driving current.

**Acknowledgements**

This work was partly supported by a Grant-in-Aid for Scientific Research on Innovative Areas "Science of Atomic Layers" from the Ministry of Education, Culture, Sports, Science and Technology (MEXT) and the Project for Developing Innovation Systems of MEXT and Grants-in-Aid for Scientific Research from the Japan Society for the Promotion of Science (JSPS) and the special fund of Institute of Industrial Science, The University of Tokyo. S. Morikawa acknowledges the JSPS Research Fellowship for Young Scientists.

**Figure captions**

Figure 1

(a) Schematic of the graphene/MoSe$_2$/Ti vertical heterostructure. (b) Fabricated device for which the layer number $N$ of MoSe$_2$ is 44. (c) Band alignment of graphene/MoSe$_2$/Ti heterostructure in OFF-state ($V_G$ = -50 V) and ON-state ($V_G$ = +50 V).

Figure 2

$I$–$V$ characteristics of graphene/MoSe$_2$/Ti vertical heterostructure with the application of $V_G$ = -50 and +50 V as measured at 300 K with MoSe$_2$ layer numbers of (a) $N$ = 14, (b) $N$ = 44, and (c) $N$ = 75. The fitting results for the OFF state $I$-$V$ curves obtained using the thermionic emission theory for determining the ideality factor $n$ are also plotted as dashed lines.

Figure 3

Temperature dependence of current density vs. $V_G$ for fixed $V_B$ = +0.1 V for graphene/MoSe$_2$/Ti vertical FETs with MoSe$_2$ layer numbers of (a) $N$ = 14 and (b) $N$ = 44. Temperature is swept at 20 K intervals. (c) Arrhenius plot at $V_G$ = +40 and -40 V. The open squares represent the data for $N$ =14; the solid circles represent the data for $N$ = 44. (d) Change in Schottky barrier height under the modulation of $V_G$.

Figure 4

(a) Change of $\varphi_{max}$ and $\varphi_{min}$ with respect to the number of MoSe$_2$ layers $N$. The dashed lines represent the calculated $\varphi_{max}$ and $\varphi_{min}$ values. (b) Current density for $V_G$ = +50 ($I_{ON}$) and -50 V ($I_{OFF}$) measured at $V_B$ = 0.5 V with respect to $N$. The dashed lines represent the calculated $I_{ON}$ and $I_{OFF}$ values from the thermionic emission theory. (c) Change in current ON-OFF ratio for different values of $N$. The dashed lines represent the calculated ON-OFF ratio from the thermionic emission theory. (d) Schematic of Schottky barrier for large and small values of $N$. $\Delta V_B$ denotes the potential energy difference between graphene with a neutral charge and the Fermi level of Ti. (e) Relationship between current modulation and barrier height modulation $\Delta\varphi$. The dashed line indicates the relationship (ON-OFF ratio) = $\exp(\Delta\varphi/k_B T)$ for $T$ = 300 K.

**References**

[1] A. K. Geim and I. V. Grigorieva, Nature **499** 419 (2013).
[2] T. Georgiou, R. Jalil, B. D. Belle, L. Britnell, R. V. Gorbachev, S. V. Morozov, Y.-J. Kim, A. Gholinia, S. J. Haigh, O. Makarovsky, L. Eaves, L. A. Ponomarenko, A. K. Geim, K. S. Novoselov, and A. Mishchenko, Nature Nanotechnol. **8** 100 (2013).
[3] W. J. Yu, Z. Li, H. Zhou, Y. Chen, Y. Wang, Y. Huang, and X. Duan, Nature Mater. **12** 246 (2013).
[4] R. Moriya, T. Yamaguchi, Y. Inoue, S. Morikawa, Y. Sata, S. Masubuchi, and T. Machida, Appl. Phys. Lett. **105** 083119 (2014).
[5] T. Yamaguchi, R. Moriya, Y. Inoue, S. Morikawa, S. Masubuchi, K. Watanabe, T. Taniguchi, and T. Machida, Appl. Phys. Lett. **105** 223109 (2014).
[6] L. Britnell, R. M. Ribeiro, A. Eckmann, R. Jalil, B. D. Belle, A. Mishchenko, Y. J. Kim, R. V. Gorbachev, T. Georgiou, S. V. Morozov, A. N. Grigorenko, A. K. Geim, C. Casiraghi, A. H. C. Neto, and K. S. Novoselov, Science **340** 1311 (2013).
[7] W. J. Yu, Y. Liu, H. Zhou, A. Yin, Z. Li, Y. Huang, and X. Duan, Nature Nanotechnol. **8** 952 (2013).
[8] A. Avsar, J. Y. Tan, T. Taychatanapat, J. Balakrishnan, G. K. W. Koon, Y. Yeo, J. Lahiri, A. Carvalho, A. S. Rodin, E. C. T. O'Farrell, G. Eda, A. H. Castro Neto, and B. Özyilmaz, Nature Commun. **5** (2014).
[9] D. Akinwande, N. Petrone, and J. Hone, Nature Commun. **5** (2014).
[10] K. F. Mak, C. Lee, J. Hone, J. Shan, and T. F. Heinz, Phys. Rev. Lett. **105** 136805 (2010).
[11] S. Tongay, J. Zhou, C. Ataca, K. Lo, T. S. Matthews, J. Li, J. C. Grossman, and J. Wu, Nano Lett. **12** 5576 (2012).
[12] H. Peelaers and C. G. Van de Walle, Phys. Rev. B **86** 241401 (2012).
[13] J. Kang, S. Tongay, J. Zhou, J. Li, and J. Wu, Appl. Phys. Lett. **102** 012111 (2013).
[14] J. S. Ross, S. Wu, H. Yu, N. J. Ghimire, A. M. Jones, G. Aivazian, J. Yan, D. G. Mandrus, D. Xiao, W. Yao, and X. Xu, Nature Commun. **4** 1474 (2013).
[15] C. R. Dean, A. F. Young, I. Meric, C. Lee, L. Wang, S. Sorgenfrei, K. Watanabe, T. Taniguchi, P. Kim, K. L. Shepard, and J. Hone, Nature Nanotechnol. **5** 722 (2010).
[16] T. Taychatanapat and P. Jarillo-Herrero, Phys. Rev. Lett. **105** 166601 (2010).
[17] S. M. Sze and K. K. Ng, *Physics of Semiconductor Devices*, 3rd ed. (Wiley, New York, 2007), pp.134.
[18] Y. Sata, R. Moriya, T. Yamaguchi, Y. Inoue, S. Morikawa, N. Yabuki, S. Masubuchi, and T. Machida, Jpn. J. Appl. Phys. **54** 04DJ04 (2015).
[19] A. Kumar and P. K. Ahluwalia, Physica B **407** 4627 (2012).
[20] C. Subhash and K. Jitendra, Semicond. Sci. Technol. **11** 1203 (1996).
[21] M. Missous and E. H. Rhoderick, J. Appl. Phys. **69** 7142 (1991).
[22] D. Sinha and J. U. Lee, Nano. Lett. **14** 4660 (2014).

Figure 1

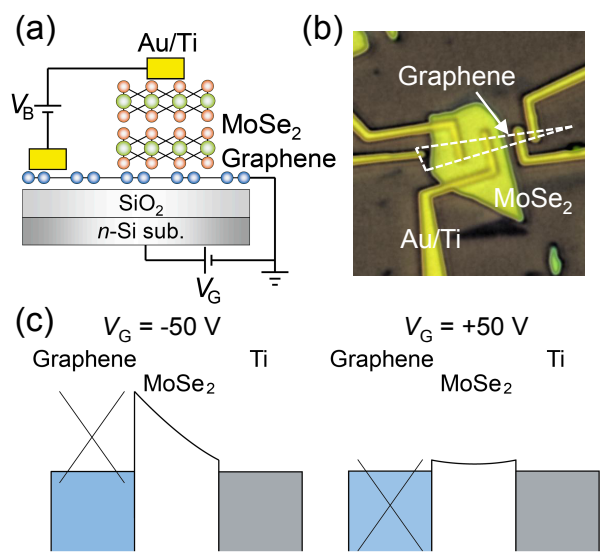

Figure 2

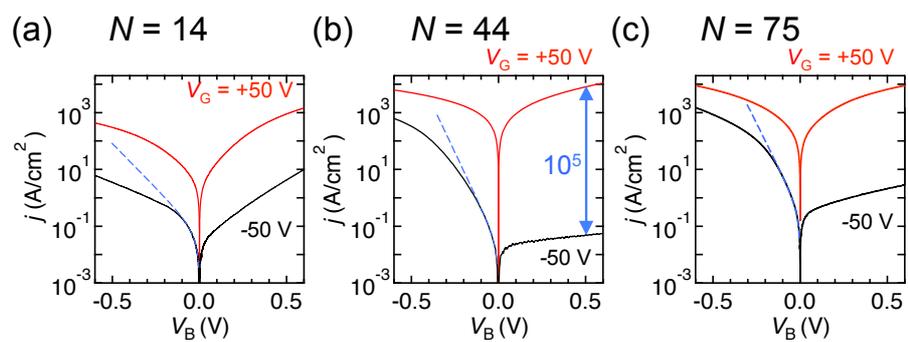

Figure 3

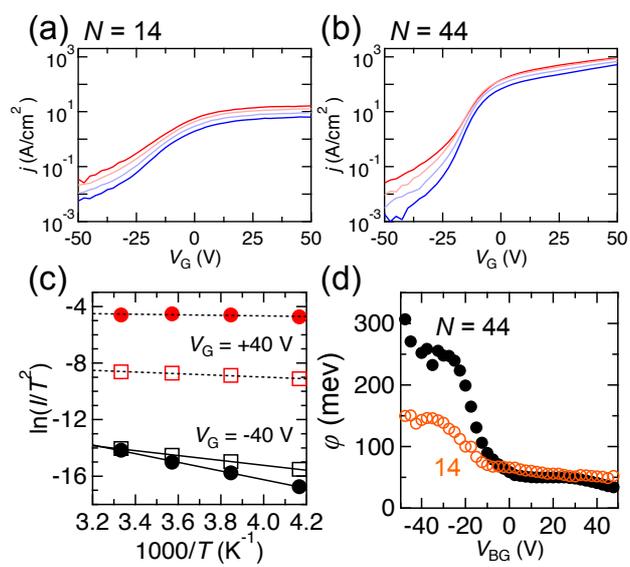

Figure 4

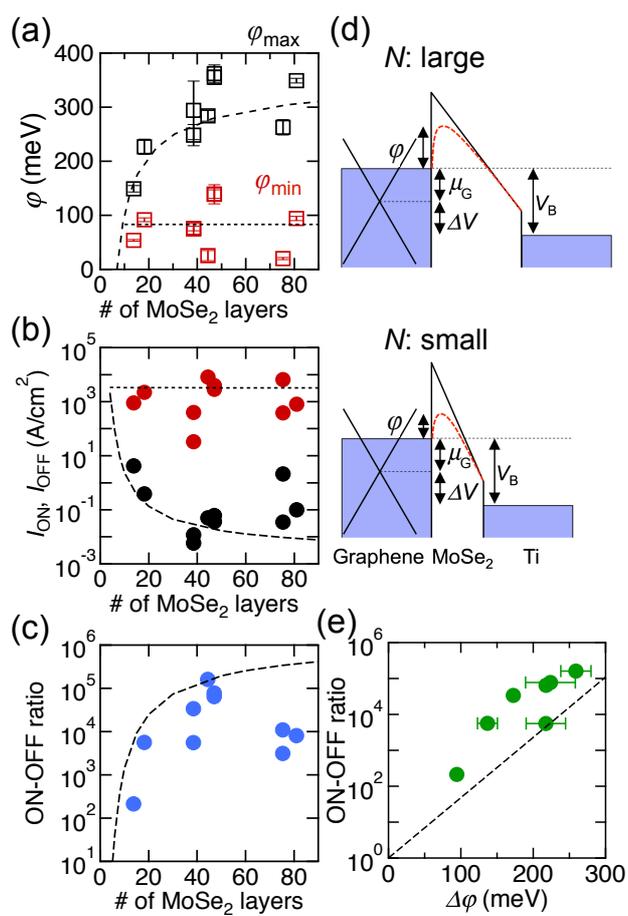